\documentclass[cameraready]{Interspeech}
\usepackage{makecell}
\usepackage{xurl}
\usepackage{IEEEtrantools}

\title{BabAR: from phoneme recognition to developmental measures\\ of young children's speech production}

\author[affiliation={1}, orcid=0000-0002-6005-9368, correspondingauthor]{Marvin}{Lavechin}
\author[affiliation={2}, orcid=0000-0003-2742-4797]{Elika}{Bergelson}
\author[affiliation={3}, orcid=0000-0002-4493-8864]{Roger}{Levy}

\address{
    $^1$ Laboratoire d'Informatique et Systèmes, Université Aix-Marseille, CNRS, France \\
    $^2$ Department of Psychology, Harvard University, USA \\
    $^3$ Department of Brain and Cognitive Sciences, Massachusetts Institute of Technology, USA
}

\email{marvinlavechin@gmail.com, elika\_bergelson@g.harvard.edu, rplevy@mit.edu}


\usepackage{comment}

\begin{document}
\bstctlcite{IEEEexample:BSTcontrol}
\maketitle

\begin{abstract}
Studying early speech development at scale requires automatic tools, yet automatic phoneme recognition, especially for young children, remains largely unsolved. Building on decades of data collection, we curate TinyVox, a corpus of more than half a million phonetically transcribed child vocalizations in English, French, Portuguese, German, and Spanish. We use TinyVox to train BabAR, a cross-linguistic phoneme recognition system for child speech. We find that pretraining the system on multilingual child-centered daylong recordings substantially outperforms alternatives, and that providing 20 seconds of surrounding audio context during fine-tuning further improves performance. Error analyses show that substitutions predominantly fall within the same broad phonetic categories, suggesting suitability for coarse-grained developmental analyses. We validate BabAR by showing that its automatic measures of speech maturity align with developmental estimates from the literature.
\end{abstract}

\section{Introduction}

\begingroup
\renewcommand\thefootnote{}
\footnotetext{%
  \hspace{-2.1em}
  \fontsize{7.3}{9}\selectfont
  To run BabAR: \url{https://github.com/MarvinLvn/BabAR}\\
  To download TinyVox: \url{https://github.com/MarvinLvn/tinyvox}\\
  Audio samples available at:\\
  \url{https://marvinlvn.github.io/projects/babar/}%
}
\endgroup

The first years of life mark a period of dramatic transformation, as most children progress from simple sounds and syllables to words and phrases, developing the ability to combine basic sound units into increasingly complex sequences~\cite{bavin2015cambridge}. Decades of research on child speech production have produced a wealth of findings with both theoretical and practical implications regarding how universal the order of learning of speech sounds is~\cite{lee2010universal}, how early babbling relates to later word production~\cite{stoel2011relationships}, how speech production is linked to various clinical conditions, and how to remediate speech difficulties~\cite{stoel1988prelinguistic}. 

Audio recordings collected in controlled or naturalistic settings are the prevailing method for studying early speech development, providing precise data on individual speech sounds and how they change over development~\cite{ambridge2013experimental}. However, this approach requires costly manual phonetic transcription, which becomes infeasible at scale. This annotation burden creates a tradeoff between depth and breadth that impedes scientific progress: longitudinal studies capturing everyday experiences provide extensive data from few children, while more controlled settings enable larger samples at the cost of sparse, brief recordings per participant (Table~\ref{tab:data} illustrates this tradeoff across the 31 corpora compiled for this study). 

Automatic speech recognition (ASR) algorithms, fueled by advances in self-supervised learning, have now reached impressive performance on adult speech, even in noisy recording conditions~\cite{Dua2023NoiseRA}. Building such systems for child speech could enable detailed phonetic tracking across thousands of hours and diverse populations, in turn accelerating research on speech acquisition and related clinical markers, opening up analyses that were previously impractical at scale.

However, child speech, especially from young infants, poses unique challenges. A newborn's vocal tract differs substantially from its adult shape: the larynx is positioned higher and the tongue fills much of the space of the oral cavity~\cite{vorperian2005development,vorperian2009anatomic}. Throughout early childhood, articulatory motor control and perceptual systems are still developing, resulting in highly variable acoustic output that differs markedly from adult speech~\cite{bavin2015cambridge,kutlu2026longitudinal}. 

Various techniques have been proposed to address the gap between child and adult speech, including vocal tract length normalization~\cite{serizel2014vocal}, data augmentation/generation~\cite{yeung2021fundamental,rolland2024improved}, multi-task learning~\cite{rumberg2021age}, and transfer learning using self-supervised models pretrained on adult speech~\cite{rolland2024introduction,Medin2024SelfSupervisedMF}. Most of this work has focused on children older than 6 years, with applications to educational technology, literacy assessment, and language learning tools~\cite{bhardwaj2022automatic,sobti2024comprehensive,li2026automated}. A notable exception, and the closest to our work, is Li et al. (2024), who trained a phoneme recognition system on the Providence corpus (including children aged 11 to 48 months) using wav2vec 2.0 pretrained on 4,300 hours of child-directed daylong recordings from American English-speaking families. However, they achieve a phoneme error rate of around 60\%~\cite{li2023foundation,li2024phoneme}, illustrating that child speech remains a formidable challenge for ASR systems. The difficulty building high-performing child ASR systems stems largely from the scarcity of publicly available annotated data, especially for non-English languages and young children. Indeed, among 27 child ASR corpora recently reviewed, only 13 include non-English data, and only one (the Providence corpus) includes children younger than 4 years old~\cite{sobti2024comprehensive}. Yet, publicly available repositories such as PhonBank contain decades of child speech recordings, laboriously collected and annotated by generations of researchers~\cite{PhonBank}. These resources remain largely untapped by the speech processing community due to the large technical debt required to aggregate disparate sources: reconciling diverse phonetic transcriptions, resolving human inconsistencies, and extracting machine-actionable segments from varied audio. 

In this work, we introduce BabAR (BABbling Automatic Recognition), a phoneme recognition system for child speech that substantially extends prior work in both scope and scale. We make the following contributions: First, we address the data barrier by curating TinyVox, a large-scale standardization of PhonBank comprising over half a million IPA-transcribed vocalizations from 560 children aged 6 months to 8 years across 5 languages (English, French, Portuguese, German, and Spanish). Second, we systematically compare multiple self-supervised models that vary in their architecture, training set size, language coverage (English-only vs. multilingual), and whether they were trained on adult-only speech or both child and adult speech. Third, we study the impact of input context duration on phoneme recognition accuracy, demonstrating that longer context windows improve performance. Fourth, we provide detailed analyses of our best-performing model.  Finally, we validate BabAR by applying it to a held-out longitudinal dataset of 44 American English-learning children recorded monthly from 6 to 18 months, and show that automatically extracted proportions of canonical vocalizations (a widely used measure of speech maturity) align with estimates from the language development literature. Together, these contributions represent a significant step toward automated phonetic analysis of early childhood speech that can scale to the longitudinal, naturalistic recordings needed to advance developmental research.

\section{Methods}

We next present our approach to recognizing phonemes in children's speech. We begin with describing the dataset, the phoneme recognition system, and the self-supervised speech representations it builds upon. We then present our context-aware fine-tuning approach. Finally, we present the baselines against which BabAR is compared and our evaluation metric. We conclude with implementation details. 

\subsection{Datasets}

\subsubsection{TinyVox (training, validation, and test sets)} 

Our dataset of IPA-transcribed children's vocalizations was created from PhonBank~\cite{PhonBank}, a large-scale database dedicated to the study of child phonology. All available data in English, French, Portuguese, German, and Spanish were downloaded and audio files were paired with their transcript (CHAT format~\cite{macwhinney2017tools}). All audio files were converted to single-channel \SI{16}{\kilo\hertz} and only transcription files with \%pho or \%xpho tiers, indicating phonetic transcriptions, were kept. Each utterance included temporal boundaries (onset and offset) and a human-annotated phonetic transcript.

\noindent \textbf{\textit{Phonetic normalization.}} The raw transcriptions contained 967 distinct phonetic categories, including base IPA symbols with various diacritical modifications (length, aspiration, voicing, nasalization, secondary articulations, etc.). However, these categories reflect heterogeneous annotation practices, with varying levels of phonetic detail across corpora. To create a more consistent cross-linguistic target inventory, we created a 57-sound set (30 consonants and 27 vowels) based on the adult phonemic inventories of all five languages.  We excluded distinctions difficult to predict acoustically (e.g., vowel length), though alternative inventories can be created using our provided script. The 967 surface variants were mapped to the 57 targets via phonological feature edit distance using panphon~\cite{mortensen2016panphon}.

\noindent \textbf{\textit{Post-processing and cleaning steps.}} We removed all utterances with extreme durations (longer than 10 seconds or shorter than 50 ms), those containing unidentified sounds (marked as 'X', 'C', 'V', 'G', 'S', 'xxx', or '*'), and those produced by children older than 8 years old. Due to the sheer scale of PhonBank, listening to all utterances to check whether they were aligned with the transcript was not feasible. We therefore implemented a two-pass sampling procedure. In the first pass, we randomly sampled 10\% of files from each corpus and listened to 10 randomly selected utterances per file to identify corpora with systematic alignment issues. This corpus-level screening flagged six problematic corpora: PhonBLA, PaidusGerman, PaidusSpanish, and Hunkeler were excluded entirely, while Davis and KernFrench were marked for file-level review. In the second pass, we listened to approximately 10 randomly sampled utterances per file in the flagged corpora and excluded files in which fewer than 8 out of 10 utterances matched their transcripts. 

\noindent \textbf{\textit{Data quality validation.}} To assess the quality of the final dataset, we randomly sampled 200 utterances. We found 139 had relatively accurate timestamps with only the transcribed child speech, while 58 included adult speech (in the vast majority of cases), other children's speech, or toy sounds due to inaccurate timestamp boundaries; 13 had overlapping speech sounds from adults, other children, or toys, and 3 were misaligned (wrong transcript).

While these may appear problematic, they are inherent to aggregating naturalistic corpora with varying recording conditions and annotation practices. Importantly, the presence of competing speech and sounds from other sources, along with transcriptions indicating the target child phonemes, provides supervised training signal for two critical capabilities: child phoneme recognition and non-target speaker suppression. Since diarization systems still produce imprecise boundaries on naturalistic recordings~\cite{lavechin2020,babyhubert2025}, training on such data prepares the downstream phoneme recognition system for real-world deployment conditions. 

\begin{figure}[htbp]
  \centering
  \includegraphics[width=.9\linewidth]{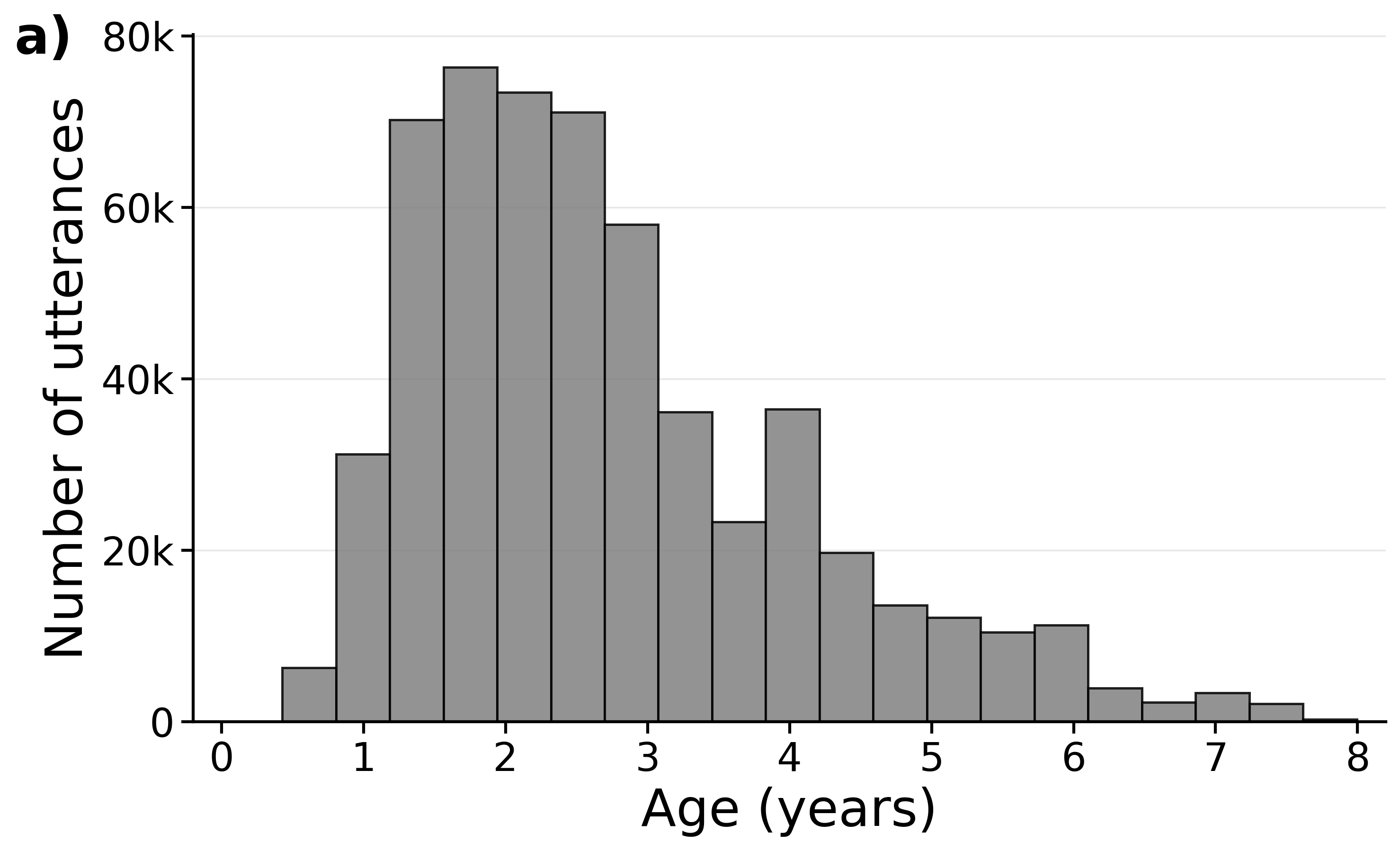}
  \includegraphics[width=.9\linewidth]{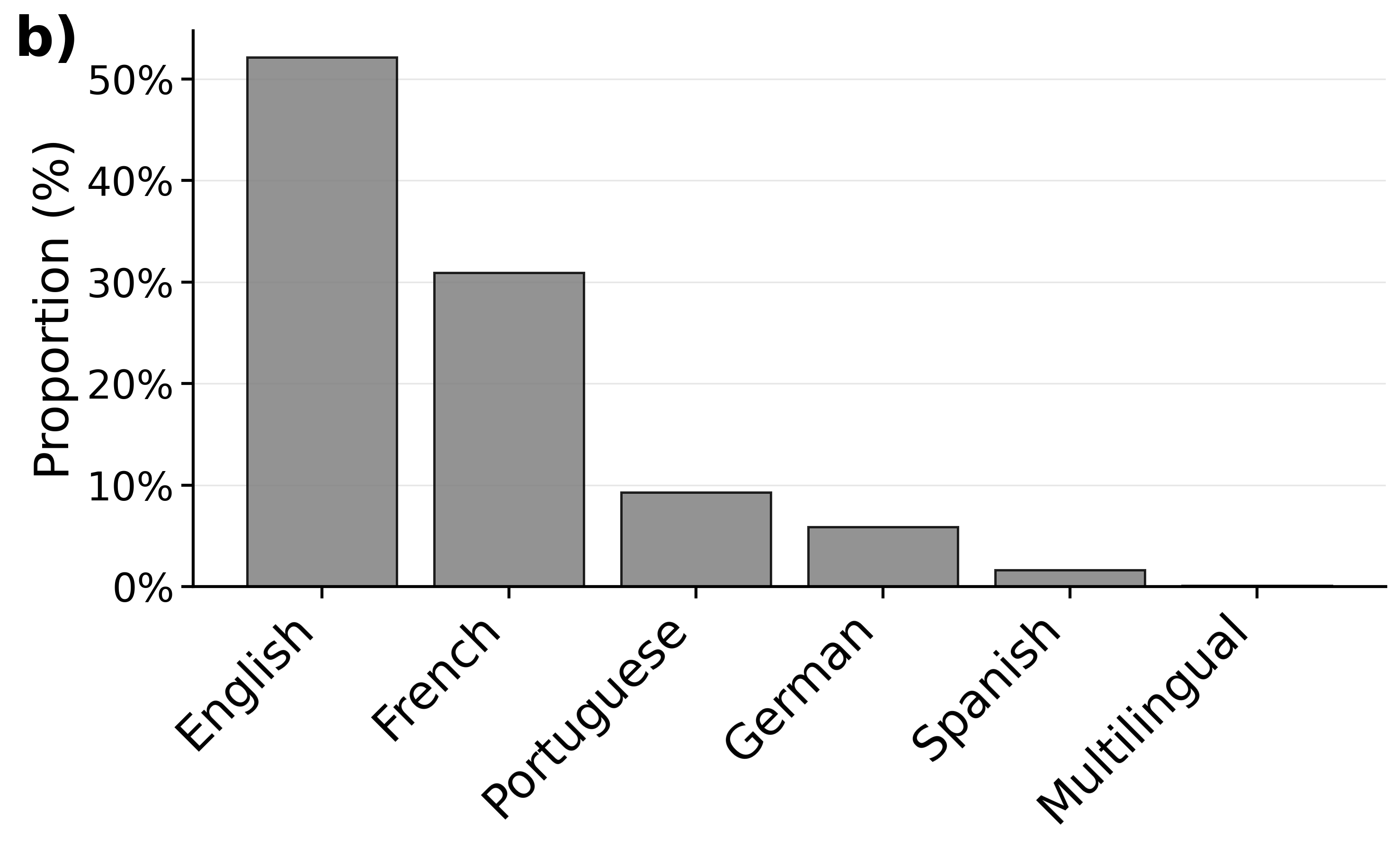}
  \caption{Age distribution (panel a) and language distribution (panel b) of phonetically transcribed utterances in TinyVox.}
  \label{fig:tinyvox}
\end{figure}

\noindent \textbf{\textit{Dataset statistics.}} Figure \ref{fig:tinyvox} shows the age and language distributions of the resulting dataset, which we named TinyVox. It contains over half a million utterances from 560 children aged 5 to 96 months (median: 29.3 months). Most of these utterances have been produced by children primarily learning English (52.1\%), followed by French (30.9\%), Portuguese (9.3\%), German (5.9\%) or Spanish (1.7\%). A minority of recordings have been collected in multilingual contexts (0.1\%). Table~\ref{tab:data} provides summary statistics for each of the 31 corpora included in TinyVox, spanning diverse recording contexts from spontaneous everyday speech to elicited production tasks (e.g., picture naming or word repetition tasks) and therapy sessions.

\begin{table*}[t]
\caption{Summary statistics for TinyVox, a dataset of child utterances transcribed into the International Phonetic Alphabet (IPA). All audio and transcript data are sourced from PhonBank~\cite{PhonBank} and have been curated to train our phoneme recognizer. Language codes: EN = English, FR = French, PT = Portuguese, DE = German, ES = Spanish. Location codes: US = United States, UK = United Kingdom, CA = Canada, LU = Luxembourg, FR = France, DE = Germany, ES = Spain, CL = Chile, PT = Portugal, BR = Brazil. L2 indicates second language learners.}
\label{tab:data}
\centering
\begin{tabular}{r c c c c c l}
\toprule
\textbf{Corpus} & \makecell{\textbf{Age mean} \\ \textbf{(mo)}} & \makecell{\textbf{Age range} \\ \textbf{(mo)}} & \makecell{\textbf{Nb. of} \\ \textbf{children}} & \makecell{\textbf{Transcribed} \\ \textbf{speech dur. (h)}} & \makecell{\textbf{Language} \\ \textbf{(Location)}} & \textbf{Activity} \\
\midrule
CCF \cite{CCF} & 30.9 & [11.0, 58.2] & 5 & 14.3 & ES (CL) & everyday activities\\ 
Cattini \cite{Cattini} & 51.7 & [35.0, 82.0] & 6 & 0.2 & FR (FR) & elicited production\\
Chiat \cite{Chiat} & 65.4 & [60.0, 68.4] & 4 & 0.5 & EN (UK) & play session\\
ChildL2 \cite{ChildL2} & 57.6 & [47.4, 75.3] & 2 & 21.2 & FR (L2) & elicited production\\
Cummings \cite{Cummings} & 56.9 & [36.6, 87.5] & 21 & 13.2 & EN (LU) & elicited production\\
Davis \cite{Davis} & 18.5 & [6.3, 36.1] & 17 & 53.4 & EN (US) & everyday activities\\
FoxBoyer \cite{FoxBoyer} & 54.1 & [27.0, 85.0] & 27 & 0.7 & DE (DE) & elicited production\\
Goad \cite{Goad} & 32.6 & [17.6, 42.6] & 2 & 3.7 & EN (CA) & play session\\
GoadRose \cite{GoadRose} & 33.4 & [12.9, 48.0] & 2 & 3.1 & FR (CA) & play session\\
Granada \cite{Granada} & 54.9 & [37.0, 71.0] & 19 & 1.3 & ES (ES) & elicited production\\
Grimm \cite{Grimm} & 20.9 & [12.2, 25.2] & 4 & 13.0 & DE (DE) & everyday activities\\
KernFrench \cite{KernFrench} & 17.7 & [7.5, 25.2] & 4 & 13.1 & FR (FR) & everyday activities\\
Lyon \cite{Lyon} & 26.8 & [11.6, 48.2] & 5 & 41.8 & FR (FR) & everyday activities\\
McAllister \cite{McAllister} & 48.5 & [45.9, 51.2] & 1 & 2.0 & EN (US) & therapy session\\
Menn \cite{Menn} & 14.4 & [12.5, 15.7] & 1 & 0.3 & EN (US) & everyday activities\\
NeumannFoxBoyer \cite{NeumannFoxBoyer} & 60.8 & [36.0, 88.0] & 26 & 0.7 & DE (DE) & elicited production\\
Paris \cite{Paris} & 33.6 & [11.6, 83.9] & 9 & 47.0 & FR (FR) & everyday activities\\
Penney \cite{Penney} & 68.2 & [59.9, 74.5] & 26 & 0.7 & EN (CA) & elicited production\\
PereiraFreitas \cite{PereiraFreitas} & 31.4 & [14.0, 55.6] & 7 & 14.3 & PT (PT) & everyday activities\\
PhonoDis \cite{PhonoDisRamalho} & 70.8 & [38.0, 93.0] & 18 & 1.4 & PT (PT) & elicited production\\
Preston \cite{Preston} & 55.1 & [48.0, 69.0] & 44 & 2.0 & EN (US) & elicited production\\
Providence \cite{Providence} & 27.6 & [11.1, 48.1] & 6 & 107.8 & EN (US) & everyday activities\\
Ramalho \cite{PhonoDisRamalho} & 57.1 & [38.5, 76.7] & 60 & 2.8 & PT (PT) & elicited production \\
Scheidnes \cite{Scheidnes} & 82.0 & [73.0, 88.8] & 33 & 0.7 & FR (L2) & elicited production\\
Seine-Marne \cite{SeineMarne} & 82.9 & [59.4, 95.9] & 26 & 0.8 & EN (L2) & elicited production\\
Stuttgart \cite{StuttgartTAKI} & 28.8 & [5.1, 91.2] & 8 & 1.6 & DE (DE) & play session\\
TAKI \cite{StuttgartTAKI} & 44.9 & [22.0, 91.2] & 4 & 1.1 & DE (DE) & play session \\
TorringtonEaton \cite{TorringtonEaton} & 60.6 & [48.7, 71.9] & 53 & 2.6 & EN (US) & elicited production\\
Twins-Brazil \cite{TwinsBrazil} & 31.3 & [14.1, 48.6] & 3 & 1.5 & PT (BR) & everyday activities\\
Vivar \cite{Vivar} & 36.5 & [24.0, 47.0] & 116 & 1.5 & ES (CL) & elicited production \\
Yamaguchi \cite{Yamaguchi} & 38.7 & [15.3, 51.9] & 1 & 6.8 & FR (FR) & everyday activities\\
\midrule
Total & 32.7 & [5.1, 95.9] & 560 & 387.7 & EN+FR+PT+DE+ES & various \\
\bottomrule
\end{tabular}
\end{table*}

\noindent \textbf{\textit{Training/validation/test split.}} The dataset was split by child rather than utterance to prevent speaker leakage between splits, targeting 80/10/10 proportions for train/validation/test. This speaker-independent split ensures we evaluate the model's ability to generalize to entirely new children at evaluation time, better reflecting real-world deployment where the system will encounter unseen speakers.

\subsubsection{The SEEDLingS corpus (held-out set)} To evaluate BabAR for developmental research, we use the SEEDLingS corpus, which consists of daylong recordings collected monthly from 44 American English-learning children aged 6 to 17 months~\cite{kalenkovich2025year,seedlingsdata}. 

\subsection{Automatic phoneme recognition} 

The phoneme recognition problem can be formulated as a sequence-to-sequence task where the input is an audio representation divided into $T$ frames $\mathbf{x} = \{x_1, \ldots, x_T\}$ at 20-millisecond intervals, and the goal is to predict the corresponding phoneme sequence $\mathbf{y} = \{y_1, \ldots, y_U\}$ where $U \leq T$ and each $y_i$ belongs to our cross-lingual 57-phoneme inventory.

Building on prior work that has shown promising performance on child speech~\cite{Medin2024SelfSupervisedMF}, we use Connectionist Temporal Classification (CTC) to handle the variable-length alignments between input frames and output phonemes. At each time step, frame $x_t$ is projected to phoneme posterior probabilities over the 57 phonemes plus a blank symbol. The CTC loss marginalizes over all valid alignments $\mathcal{A}_{\mathbf{x},\mathbf{y}}$:

\begin{equation*}
\mathcal{L}_{CTC} = -\log p(\mathbf{y}|\mathbf{x}) = -\log \sum_{A \in \mathcal{A}_{\mathbf{x},\mathbf{y}}} \prod_{t=1}^{T} p_t(a_t|\mathbf{x})
\end{equation*}

\noindent where $a_t$ denotes the symbol (phoneme or blank) at time step $t$ in alignment path $A$, and $p_t(a_t|\mathbf{x})$  is the probability predicted by the model for that symbol at time $t$.

The frame representations $\mathbf{x}$ are extracted from self-supervised models, which we describe next.

\subsection{Self-supervised models} 

\begin{table}[htbp]
\caption{Self-supervised models evaluated in this study and their pretraining set characteristics. Dur. = total duration in hours; Adult = presence of adult speech; Child = presence of child speech; Multi. = multilingual speech; Nat. = naturalistic recordings capturing spontaneous speech and other sounds as they occur in everyday life. Checkmarks indicate the presence of each characteristic.}
\label{tab:models}
\centering
\setlength{\tabcolsep}{3pt}
\begin{tabular}{r c c c c c}
\toprule
& \multicolumn{5}{c}{\textbf{Pretraining set characteristics}} \\
\textbf{System} & \textbf{Dur. (h)} &\textbf{Adult} & \textbf{Child} & \textbf{Multi.} & \textbf{Nat.}\\
\midrule
W2V2 \cite{wav2vec22020} & 960 & \checkmark &  &  \\
HuBERT \cite{hubert2021} & 960 & \checkmark & & \\
WavLM \cite{chen2022wavlm} & 960 & \checkmark & & \\
W2V2 XLSR \cite{xlsr2021} & 53,000 & \checkmark & & \checkmark \\
W2V2 LL4300 \cite{li2023foundation} & 4,300 & \checkmark & \checkmark &  & \checkmark \\
BabyHuBERT \cite{babyhubert2025} & 13,000 &\checkmark & \checkmark & \checkmark & \checkmark \\
\bottomrule
\end{tabular}
\end{table}

We evaluate six self-supervised models that vary in their training data characteristics (Table \ref{tab:models}) and their architecture: wav2vec 2.0 (hereinafter referred to as W2V2), HuBERT, or WavLM. All three architectures use a stack of convolutional layers to encode raw audio waveforms into features, followed by transformer layers that process these features into context-dependent representations. W2V2 uses a contrastive learning objective where the model learns to distinguish true future representations from distractors~\cite{wav2vec22020}. HuBERT employs a masked-prediction objective in which the model predicts quantized representations of masked audio segments~\cite{hubert2021}. WavLM extends the masked prediction framework with an additional denoising objective~\cite{chen2022wavlm}. The base versions of W2V2, HuBERT, and WavLM were trained on 960 hours of English audiobook recordings (LibriSpeech~\cite{panayotov2015librispeech}). W2V2 XLSR adapts the W2V2 large architecture to 53,000 hours of multilingual adult speech~\cite{xlsr2021}. W2V2 LL4300 uses the base architecture but was trained on 4,300 hours of child-centered daylong recordings from American English-speaking families, including both child and adult speech~\cite{li2023foundation}. BabyHuBERT adapts the HuBERT base architecture and was trained on 13,000 hours of multilingual child-centered daylong recordings containing both child and adult speech~\cite{babyhubert2025}. To the best of our knowledge, W2V2 LL4300 and BabyHuBERT are the only two self-supervised models pretrained on child-centered daylong recordings. Such pretraining is likely beneficial for child phoneme recognition, providing exposure to child speech, overlapping speech, background noise, and variable acoustics.

\subsection{Context-aware fine-tuning with extended audio}

Beyond model pretraining, we hypothesize that providing extended audio context during fine-tuning can further help the model adapt to challenging recording conditions, potentially improving phoneme recognition performance.

Following recent work on context-aware ASR training~\cite{schwarz21_interspeech,howmuch2024,altinok2025mind}, we explore whether similar techniques can improve child phoneme recognition. For an utterance with annotated boundaries $[t_{\text{start}}, t_{\text{end}}]$, we extract audio from the extended window $[t_{\text{start}} - c/2, t_{\text{end}} + c/2]$ where $c$ represents the context duration in seconds. The full extended window is fed to the encoder to produce frame-level representations. However, the CTC loss is computed only on the frames corresponding to the target utterance. Similarly, at inference time, we extract predictions only from the target utterance frames, discarding predictions from the surrounding context. In other words, the transformer layers have access to all the context provided, while logits from non-target-speech segments (potentially unannotated) do not contaminate either the loss computation or the final predictions.

We systematically evaluate the impact of context duration by measuring performance with $c \in \{0, 5, 10, 15, 20, 25, 30\}$ seconds, where $c = 0$ corresponds to the baseline model without context (i.e., the model receives only the target utterances). Unless otherwise specified, all reported results use models trained without context ($c = 0$).

\subsection{Baselines}

We evaluate BabAR's performance against two state-of-the-art phoneme recognition systems. W2V2Phoneme~\cite{xu22b_interspeech} is a phoneme recognizer built by fine-tuning the W2V2-XLSR architecture on labeled phoneme transcriptions from multiple languages. ZIPA~\cite{zhu-etal-2025-zipa} is a Zipformer architecture trained from scratch on 17k hours of multilingual IPA-transcribed speech. Both systems are used off-the-shelf with their publicly available checkpoints\footnote{Available at \url{https://huggingface.co/facebook/wav2vec2-lv-60-espeak-cv-ft} and \url{https://huggingface.co/anyspeech/zipa-large-crctc-ns-800k}, respectively.}. To ensure performance metrics are consistent across systems, we map the predicted phonetic categories to our 57 target categories via phonological feature edit distance using panphon~\cite{mortensen2016panphon}. While neither system was trained on child speech, existing child phoneme recognition systems are typically monolingual and thus unable to predict our full cross-linguistic phoneme inventory. These two universal phone recognizers are therefore the most appropriate baselines for our cross-linguistic setting. However, our evaluation task is twofold: models must both extract the target child's speech from segments with noisy boundaries that may contain adult speech or other competing signals, and transcribe it into phonemes. Since neither baseline was designed to handle such conditions, we expect their performance to be substantially degraded on TinyVox.

\subsection{Evaluation metric}

To evaluate our models, we use the commonly used phoneme error rate (PER) defined as:

\begin{equation*}
\text{PER} = 100 \times \dfrac{I + D + S}{N} 
\end{equation*}

\noindent where $I$, $D$, and $S$ count the insertions, deletions, and substitutions needed to align the predicted sequence to the reference, and $N$ is the total number of phonemes in the reference.

\subsection{Implementation details}

\noindent \textit{\textbf{Training.}} We fine-tune the self-supervised models by adding a two-layer feed-forward prediction head on top of the encoder: a hidden layer of 384 dimensions followed by a final projection to the 57-phoneme vocabulary, with rectified linear unit (ReLU) activation and dropout (p=0.1). All self-supervised models are initialized from their publicly available pretrained checkpoints. 

Following~\cite{Medin2024SelfSupervisedMF,chen2022wavlm}, we freeze the convolutional layers throughout training while keeping the transformer layers and the prediction head trainable and use the AdamW optimizer with a learning rate of $10{-5}$ and a weight decay of $10^{-2}$. As in~\cite{chen2022wavlm}, we use a tri-stage learning rate scheduler with a linear warm-up from 0 to the peak learning rate over $10\%$ of training steps ($10,000$ steps), a constant learning rate for $40\%$ of steps ($40,000$ steps), followed by a linear decay to 0 over the remaining $50\%$ steps. Training is conducted for a maximum of $100,000$ steps (approximately $21$ epochs). We use mixed-precision training (FP16) and gradient accumulation with a batch size of 32 and 2 accumulation steps, yielding an effective batch size of 64. Models are trained on a single NVIDIA V100 GPU (32GB) for approximately 5 days, and each configuration is trained with 5 random seeds.

\noindent \textbf{\textit{Validation and test.}} We validate after each training epoch and select the best checkpoint based on validation PER. During inference (validation and test), we use greedy decoding by selecting the most probable phoneme at each time step and collapsing repeated predictions, as beam search with N-gram language models yielded no performance gains.

\section{Results}

We first compare the performance of six self-supervised models fine-tuned on TinyVox, then examine the effect of providing extended audio context during training. We then compare the performance of our best-performing system, BabAR, against two baselines, before analyzing its error types. Finally, we close the loop from model development to real-world application by using BabAR to measure speech production development on a held-out set of naturalistic child-centered recordings unseen during training, validation, and evaluation.

\subsection{Which self-supervised model performs best?}

\begin{figure}[htbp]
  \centering
  \includegraphics[width=\linewidth]{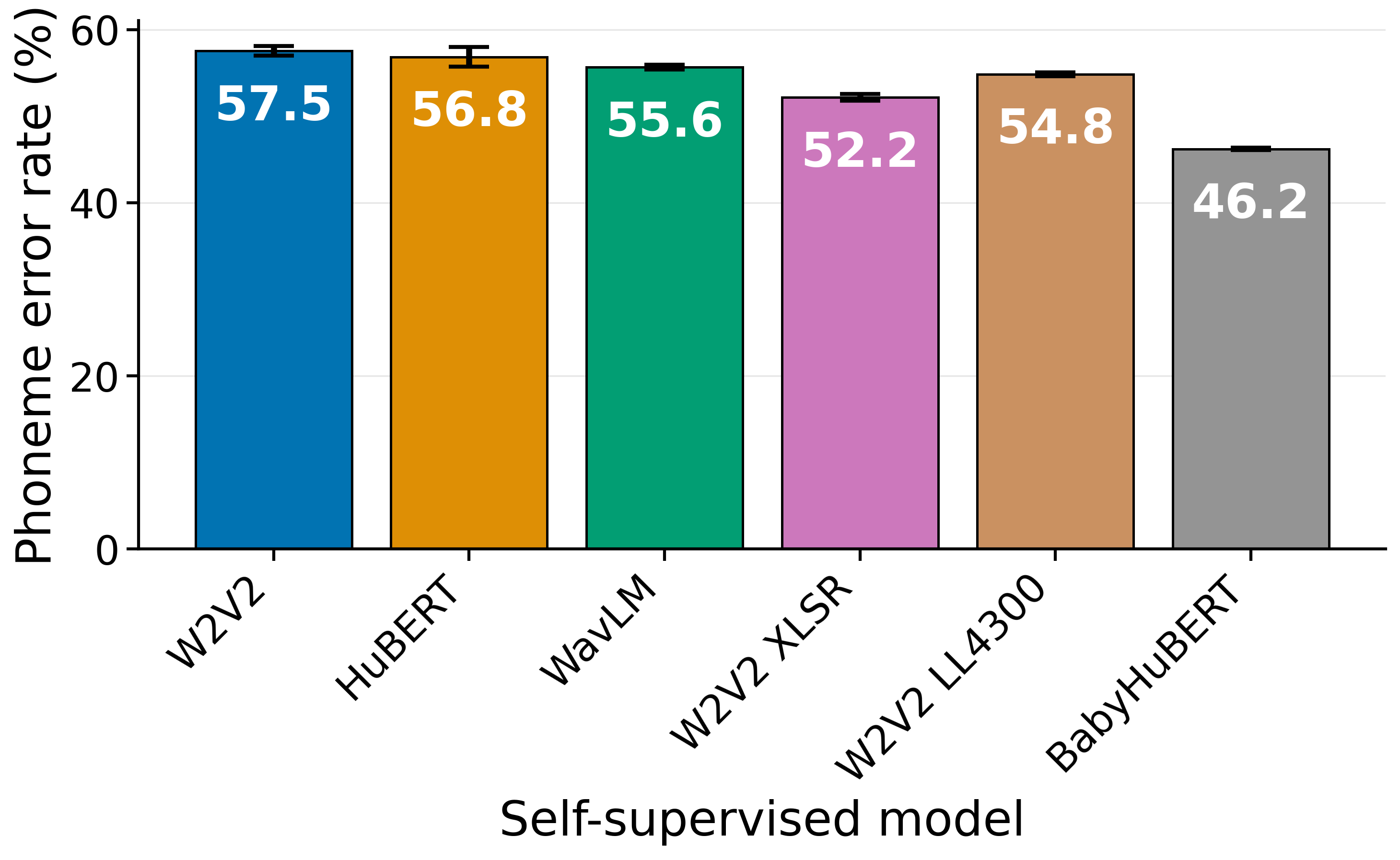}
  \caption{Validation phoneme error rate (\%, lower is better) for different self-supervised models fine-tuned on TinyVox. Means and standard deviations are computed across 5 training seeds.}
  \label{fig:ssl_choice}
\end{figure}

Before we turn to comparing the six self-supervised models in Figure~\ref{fig:ssl_choice}, we note that, while we can identify which model performs best, attributing performance differences to specific aspects of the pretraining is challenging given the many confounding variables. Models differ simultaneously in pretraining data size, language coverage, architectural choices, and implementation details (see Table~\ref{tab:models}). 

With this caveat in mind, we observe several patterns. First, among the three models pretrained on LibriSpeech (W2V2, HuBERT, WavLM), performance is relatively similar, with WavLM achieving slightly better performance ($55.6 \pm 0.27\%$ PER). Given that all three models have been pretrained on the same dataset and share similar architectures, we hypothesize that WavLM's denoising objective during pretraining may provide an advantage for TinyVox, which contains naturalistic recordings with overlapping speech and variable acoustic conditions. This aligns with findings from Medin et al. (2024)~\cite{Medin2024SelfSupervisedMF}.

Second, W2V2 XLSR, trained on multilingual adult speech, outperforms its monolingual counterpart W2V2 ($52.2 \pm 0.38\%$ vs $57.5 \pm 0.57$\% PER). This $5.3\%$ absolute difference likely stems from multiple factors: a higher number of parameters (large vs. base architecture), substantially more pretraining data (53K vs. 960 hours), and multi vs. monolingual coverage.

Third, both models pretrained on child-centered recordings show contrasting results. Surprisingly, W2V2 LL4300 (4.3K hours, English-only, base architecture) achieves $54.8 \pm 0.23\%$ PER, performing similarly to LibriSpeech models, despite substantially more pretraining data and exposure to child speech. However, this aligns with Li et al.~\cite{li2024phoneme} who found that fine-tuning W2V2 LL4300h directly on Providence (included in TinyVox) caused training to diverge. In contrast, BabyHuBERT (13K hours, multilingual, base architecture) achieves the best performance at $46.2 \pm 0.15\%$ PER: a $6.0\%$ absolute improvement over W2V2 XLSR despite using a base rather than a large architecture and four times less pretraining data. 

Overall, these results suggest that pretraining on naturalistic, multilingual, child-centered daylong recordings is beneficial for child phoneme recognition. Given BabyHuBERT's superior performance, we use it for all subsequent experiments.

\subsection{How much context helps?}

\begin{figure}[htbp]
  \centering
  \includegraphics[width=\linewidth]{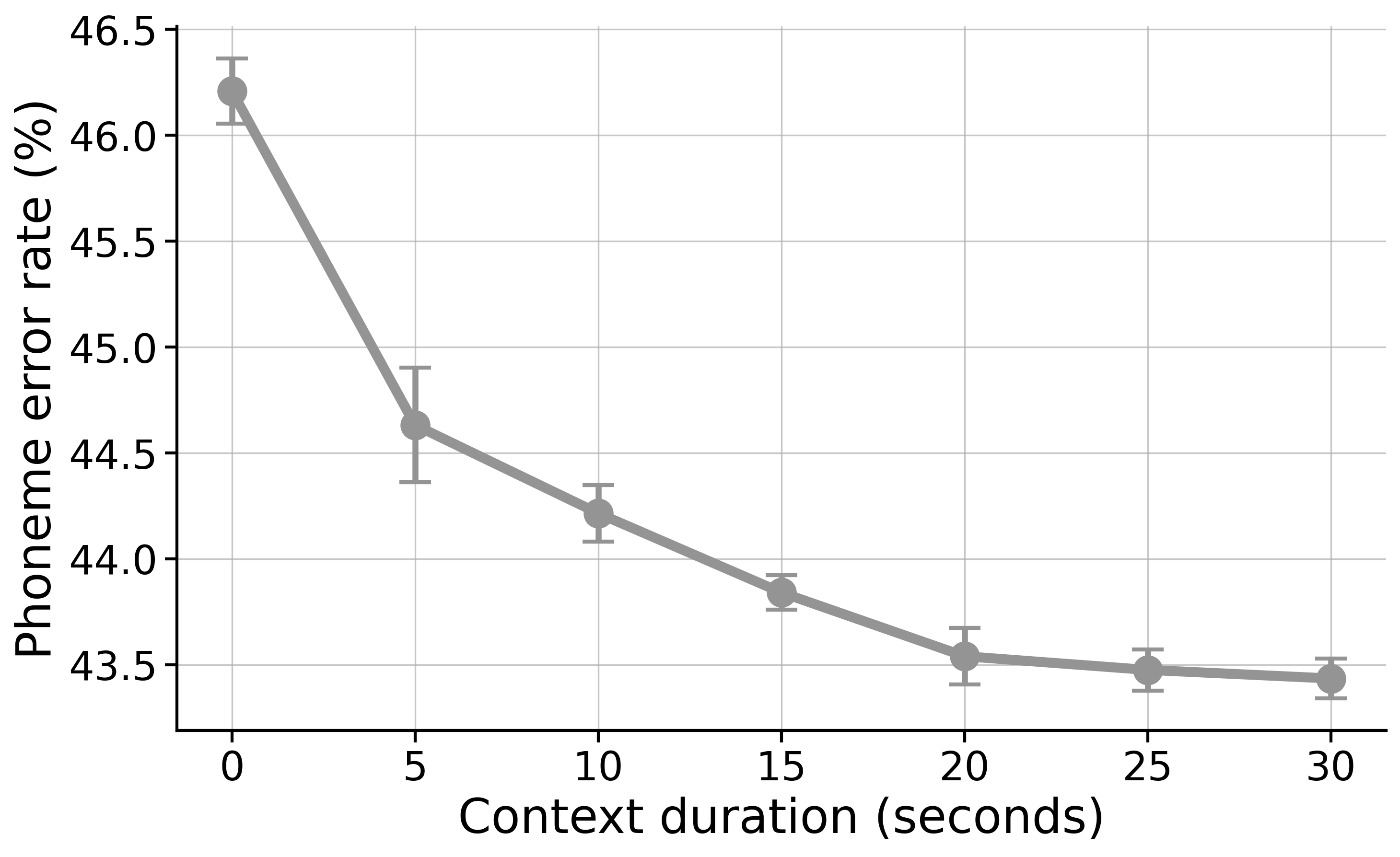}
  \caption{Validation phoneme error rate (\%, lower is better) for BabAR (BabyHuBERT fine-tuned on TinyVox) as a function of context duration $c$. $c = 0$ corresponds to the model receiving only the target child speech utterances (using human-annotated boundaries). Error bars represent the standard deviation across 5 training seeds. (\textit{N.B.}: truncated y-axis)}
  \label{fig:context}
\end{figure}

Figure~\ref{fig:context} shows the impact of context duration on phoneme recognition performance. We observe consistent improvements as context increases from 0 to 20 seconds, with validation PER decreasing from $46.2 \pm 0.15\%$ (no context) to $43.5 \pm 0.13\%$ (20 seconds), representing a $2.7\%$ absolute difference. 

However, the rate of improvement diminishes with increasing context: the largest gains occur in the first 10 seconds ($2.0\%$ absolute reduction), while extending from 10 to 20 seconds yields a smaller improvement ($0.7\%$ absolute). Beyond 20 seconds, performance plateaus, with 25 and 30 seconds of context providing no additional benefit.

The extended context likely helps the model in multiple ways: discriminating the target child's speech from competing sources (adults, other children, or environmental sounds), adapting to that child's voice characteristics and articulatory patterns through their other nearby vocalizations, and leveraging surrounding linguistic context such as adult prompts (e.g., "say daddy"). However, these benefits are largely captured within a 20-second window. For the remaining analyses, we use the best-performing configuration with $c=20$ seconds.

\subsection{What types of errors does BabAR make?}

We now analyze the errors made by BabAR on the test set, first comparing performance to baselines, then examining how error rates vary across languages, and finally investigating whether substitution errors tend to remain within or fall across broad phonetic categories.

\subsubsection{Insertion, deletion, and substitution rates}

\begin{table}[htbp]
\caption{Insertion (I), deletion (D), substitution (S) rates (\%) and phoneme error rate (PER, \%) computed on the TinyVox test set for our baseline systems against BabAR. For all metrics, lower is better. The best-performing system is indicated in bold.}
\label{tab:test_perf}
\centering
\setlength{\tabcolsep}{4.9pt}
\begin{tabular}{r c c c c}
\toprule
& \multicolumn{4}{c}{\textbf{Performance metrics}} \\
\textbf{System} & \textbf{I (\%)} & \textbf{D (\%)} & \textbf{S (\%)} & \textbf{PER (\%)} \\
\midrule
W2V2Phoneme~\cite{xu22b_interspeech} & 59.5 & 18.6 & 51.8 & 129.9\\
ZIPA~\cite{zhu-etal-2025-zipa} & 60.1 & 18.0 & 46.2 & 124.3\\
BabAR (ours) & \textbf{4.9} & \textbf{15.8} & \textbf{21.4} & \textbf{42.1}\\
\bottomrule
\end{tabular}
\end{table}

As shown in Table~\ref{tab:test_perf}, both baseline systems exhibit extremely high error rates on TinyVox, with PERs exceeding $120\%$. This is primarily driven by high insertion rates ($59.5\%$ for W2V2Phoneme, $60.1\%$ for ZIPA-large). These high insertion rates largely reflect the noisy segmentation boundaries in TinyVox: utterances often contain untranscribed competing signals such as adult speech, which these models readily transcribe as additional phonemes.  Substitution rates are also high ($51.8\%$ and $46.2\%$, respectively), suggesting that these systems also struggle to correctly identify the phonemes produced by the target child. These results are not surprising, as both systems were trained on adult speech and are neither adapted to the acoustic characteristics of child vocalizations nor robust to the presence of untranscribed competing signals.

By contrast, BabAR reduces PER by over $80$ percentage points absolute, with the most dramatic improvement in insertion rate (from $\sim60\%$ to $4.9\%$), indicating that domain-specific fine-tuning enables the model to focus on the target child's speech while ignoring competing signals. Substitutions nonetheless remain the dominant source of errors at $21.4\%$, underscoring the difficulty of mapping child speech to canonical phoneme targets, while deletions ($15.8\%$) suggest that BabAR tends to miss speech sounds rather than predict too many. 

For context, current phoneme recognition systems achieve PERs below $10\%$ on adult read speech recorded in clean conditions (e.g.,~\cite{chen2022wavlm}). The substantially higher error rates we observe reflect both the challenging recording conditions of naturalistic child speech and the difficulty of mapping the productions of still-developing vocal tracts onto phonetic categories that were initially designed to describe adult speech sounds. We contextualize these error rates in light of human inter-annotator agreement studies in the Discussion (Section~\ref{sec:discussion}).

\subsubsection{Phoneme error rate across languages} We observe different test PER across languages with $32.7\%$ PER for English (computed from 28.7 hours of speech), $52.4\%$ for French (23.0 hours), $32.9$ for Portuguese (0.5 hour), $37.5\%$ for German (0.7 hour), $26.5\%$ for Spanish (0.5 hour). However, these differences should be interpreted with caution, as several confounding factors vary across our test languages, including the age distribution of the children, the types of recorded activities, recording conditions, annotation quality, and number of phonemes in the target language. Evaluating the cross-linguistic performance of BabAR is an important direction, but doing so reliably would require a more carefully balanced test set, which we leave for future work.

\subsubsection{Substitution rates across phonetic categories} Rather than requiring exact phoneme-level accuracy, many downstream developmental analyses operate at coarser levels of granularity, for instance, tracking the ratio of consonants to vowels across age, or measuring how consonants' place or manner distributions shifts over the first years of life. For such analyses, a substitution between two stops (e.g., /t/ $\rightarrow$ /k/) is far less consequential than a substitution across manner classes (e.g., a stop predicted as a fricative), since only the latter would distort the derived developmental metrics. This motivates examining whether substitution errors tend to remain within or fall across broad phonetic categories. 

\begin{figure*}[htbp]
  \centering
  \includegraphics[width=0.46\textwidth]{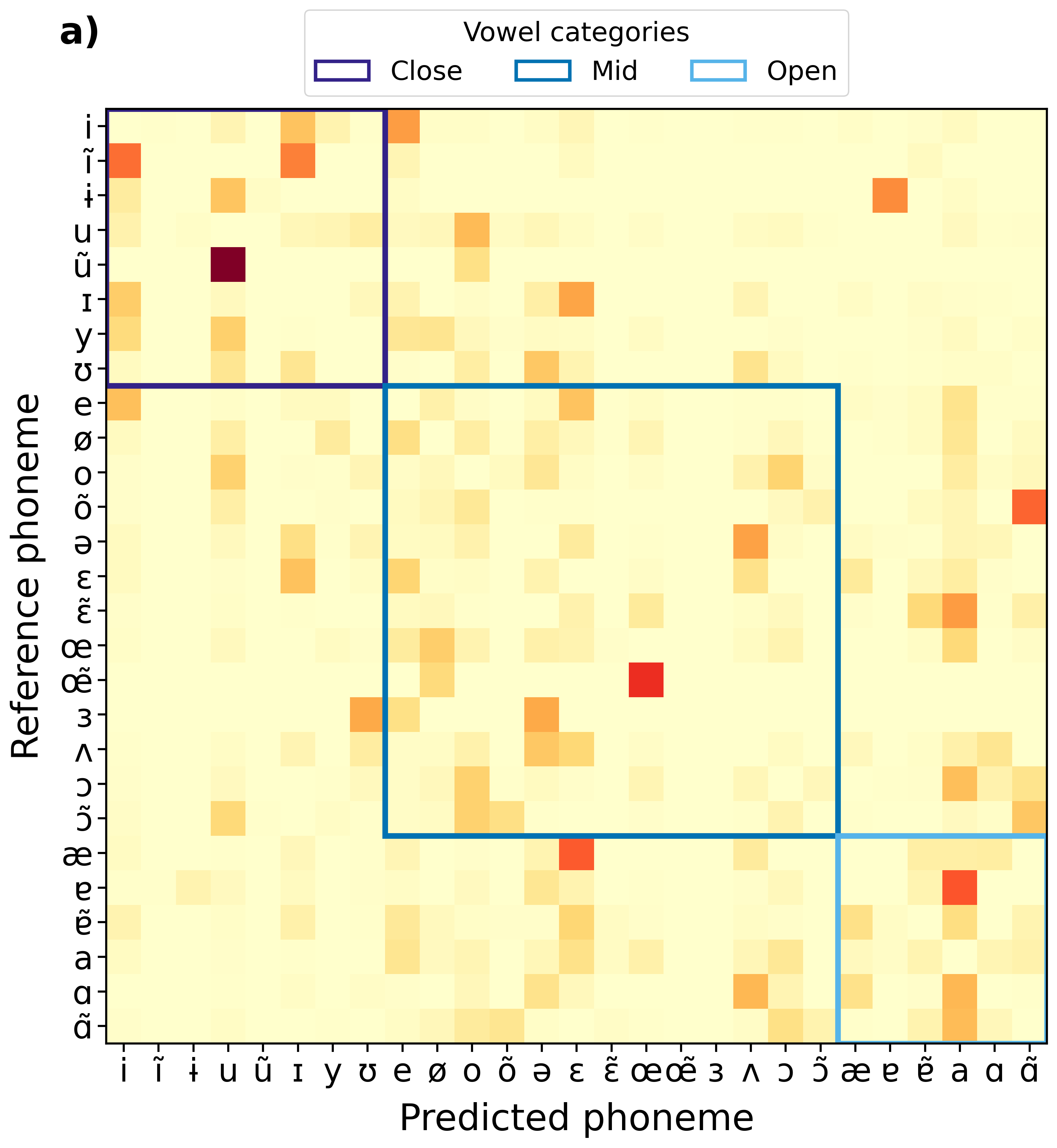}%
  \hfill
  \includegraphics[width=0.50\textwidth]{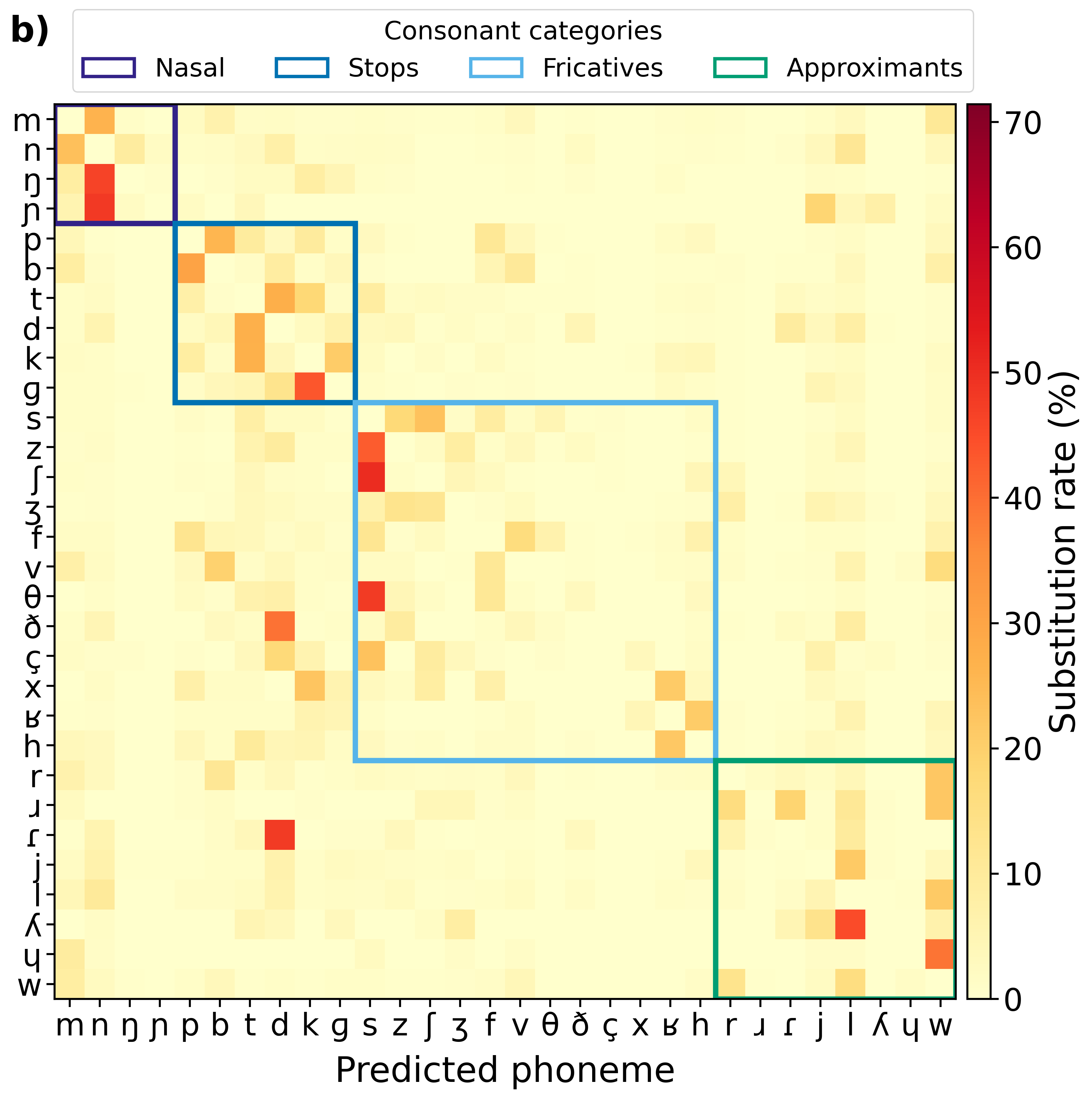}
  \caption{Substitution matrices for vowels (panel a) and consonants (panel b) indicating which substitution errors BabAR makes. The darker the cell, the higher the substitution rate. All numbers are computed on the test set of TinyVox. Substitutions were more likely within vowel/consonant categories (in each outlined square) than across them.}
  \label{fig:confusion}
\end{figure*}

To do so, we first compute a substitution matrix, where each cell $(i,j)$ reports the rate at which reference phoneme $i$ is substituted by predicted phoneme $j$, normalized by the total number of substitution errors for phoneme $i$. In other words, this analysis asks: when BabAR makes a substitution error on a given phoneme, which phoneme does it predict instead? 

Since phoneme substitutions rarely cross vowel/consonant boundaries (consonants: $64.8\%$ correct, $13.6\%$ substituted by 
a consonant, $4.0\%$ by a vowel, $17.6\%$ deleted; vowels: 
$60.7\%$ correct, $22.2\%$ substituted by a vowel, $3.3\%$ by 
a consonant, $13.9\%$ deleted), analyzing vowels and consonants separately gives a clearer picture of error patterns within each category; we therefore report results separately in Figure~\ref{fig:confusion}.

For vowels (Figure~\ref{fig:confusion}a), we highlight several patterns at increasing levels of granularity. At the level of individual sounds, consider /\textipa{\~I}/: when substituted, $40\%$ of errors map to /\textipa{i}/ and $37.5\%$ to /\textipa{I}/, differing from /\textipa{\~I}/ in tenseness or nasalization. Similarly, when /\textipa{\~u}/ is substituted, $71.4\%$ of errors map to /\textipa{u}/, both close back rounded vowels differing in nasalization. These examples illustrate a broader pattern: when nasalized vowels are substituted, the most frequent substitute tends to be their oral counterpart (e.g., /\textipa{\~I}/ $\rightarrow$ /\textipa{I}/, /\textipa{\~u}/ $\rightarrow$ /\textipa{u}/, /\textipa{\~{\oe}}/ $\rightarrow$ /\textipa{\oe}/), suggesting that the nasalization contrast is a primary source of confusion for the model. At the level of vowel height, we compute for each category the proportion of substitution errors that remain within the same height category (i.e., within the colored blocks in Figure~\ref{fig:confusion}a) versus crossing into another. For close vowels, $52.2\%$ of substitution errors map to another close vowel, $38.4\%$ to a mid vowel, and only $9.3\%$ to an open vowel. Mid vowels show a similar within-category rate ($52.6\%$), but with more symmetric spillover toward both close ($19.9\%$) and open ($27.4\%$) vowels. Open vowels have the lowest within-category rate: only $37.9\%$ of substitution errors map to another open vowel, while $54.2\%$ map to mid vowels and $7.9\%$ to close vowels.

For consonants (Figure~\ref{fig:confusion}b), we observe a similar structure. At the level of individual sounds, consider /\textipa{N}/: when it is substituted, $46.5\%$ of errors map to /\textipa{n}/, both nasal consonants differing only in place of articulation.  Similarly, when /\textipa{t}/ is substituted, $27.6\%$ of errors map to /\textipa{d}/ and $17.8\%$ to /\textipa{k}/, both stops different in voicing or place. More broadly, substitution errors tend to stay within the same manner of articulation. Quantifying this at the category level (i.e., within the colored blocks in Figure~\ref{fig:confusion}b): $63.1\%$ of substitution errors on stops map to another stop, $55.8\%$ of errors on nasal map to another nasal, $55.0\%$ of errors on approximants map to another approximant, and $51.1\%$ of errors on fricatives map to another fricative.

Thus, although BabAR's test PER of $42.1\%$ reflects frequent phoneme-level errors, substitutions tend to remain within broad phonetic categories, suggesting it may be more reliable than its PER alone implies for downstream analyses at coarser levels of granularity, such as consonant-to-vowel ratios or manner-class distributions.

\subsection{Can BabAR measure speech production development?}

The preceding analyses evaluate BabAR's recognition performance on test data drawn from the same corpora used for training, but our initial goal was to measure the child's speech production at scale, without manual annotation. A stronger test, then, is whether BabAR can recover known developmental trends from unseen naturalistic recordings. To this end, we built a fully automatic pipeline to transcribe the child's production and apply it to the SEEDLingS corpus, containing monthly recordings from 44 American English-learning children. We first use VTC 2.0 (Voice Type Classification,~\cite{babyhubert2025}) to detect vocalizations from the child wearing the recording device, then run BabAR on the detected vocalizations to obtain phoneme transcriptions. From these transcriptions, we compute the proportion of utterances containing at least one consonant-vowel (CV) or vowel-consonant (VC) transition, a well-established measure of speech maturity known as canonical proportion. We compare these automatically derived estimates to the developmental trajectory reported in Cychosz \& Long (2025), a meta-analysis of $43$ studies including $1,291$ infants.

\begin{figure}[htbp]
  \centering
  \includegraphics[width=\linewidth]{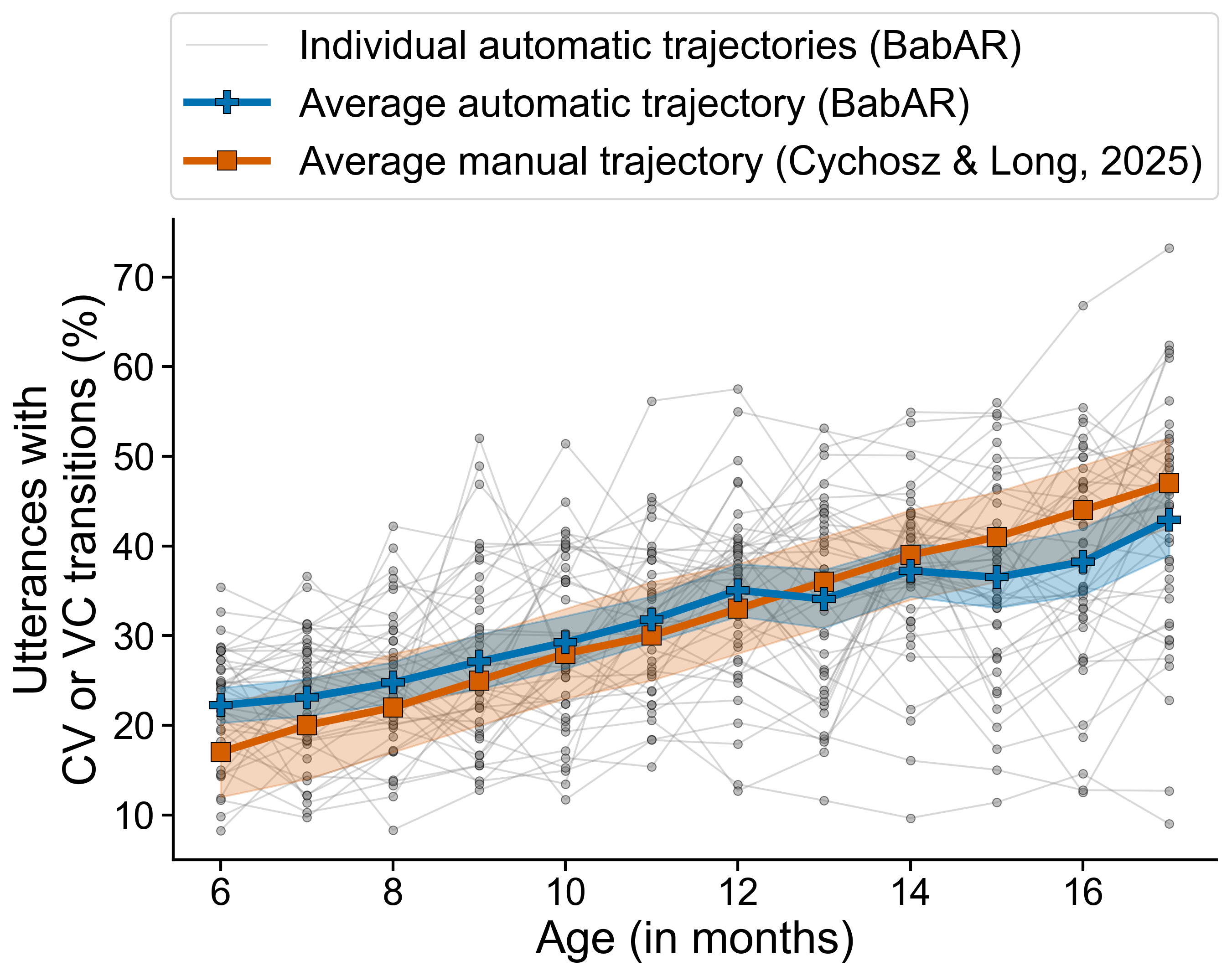}
  \caption{Proportion of utterances with consonant-vowel (CV) or vowel-consonant (VC) transitions as a function of age (in months). Gray lines show individual trajectories computed by BabAR for 44 American English-learning children from SEEDLingS, and the blue curve shows the corresponding average. The orange curve shows the average trajectory derived from manual annotation from a meta-analysis by Cychosz \& Long (2025). Shaded areas indicate $95\%$ confidence intervals.}
  \label{fig:cp_utt}
\end{figure}

Figure~\ref{fig:cp_utt} shows the results. Both the average automatic and manual trajectories exhibit a clear increasing trend from 6 to 17 months. Crucially, BabAR's automatic trajectory falls within the 95\% confidence interval of the manual trajectory across the entire age range, indicating that our fully automatic pipeline produces group-level canonical proportion estimates consistent with what we know about American English-learning children from the literature.

These results demonstrate that our fully automatic pipeline, from vocalization detection to phoneme transcription, recovers an established group-level developmental trend on this corpus of 44 American English-learning children.

\section{Discussion}
\label{sec:discussion}

BabAR represents a first step toward automated phoneme-level transcription of young children's speech at scale, opening new possibilities for large-scale screening for speech and language delays, cross-linguistic comparisons of phonological development, and investigation of how early vocal patterns relate to later language outcomes — analyses that have until now relied on manual transcription, limiting them to small samples. Recent work has begun to address this gap with utterance-level vocalization classifiers~\cite{zhang25r_interspeech} and entropy-based developmental measures~\cite{sy2023measuring}, but BabAR is, to our knowledge, the first to provide phoneme transcriptions of young children's speech.

Our developmental validation provides an encouraging proof of concept for this. When applied to a held-out set of naturalistic recordings, BabAR's derived canonical proportion estimates fall within the confidence interval of a large-scale meta-analysis based on manual annotations, without any manual annotation of the target recordings. This is likely because canonical proportion operates at a coarse level of granularity and is computed by aggregating over hundreds of vocalizations per child and age point, both of which help average out phoneme-level errors. Whether finer-grained measures, such as consonant inventory size, manner class distributions, or phonological complexity indices, can be reliably extracted remains to be tested. This first validation is also conducted at the group level: the average automatic trajectory aligns with the literature, but this does not guarantee that individual trajectories are sufficiently accurate to detect meaningful differences between children, a distinction that matters for clinical applications, where the goal is often to identify children whose vocal development falls below age-expected norms. Addressing these questions would require comparing BabAR's estimates against manual transcriptions for the same children across diverse populations.

Building phoneme recognition systems for child speech requires acknowledging two challenges inherent to this domain. The first is the ground truth itself. Phonetic transcription is not an objective record of what was produced by the child, but a subjective perceptual judgment, and human transcribers disagree substantially when transcribing child speech. Although reported inter-annotator agreement scores range widely, from $51\%$ to $97\%$, there is consensus that several factors influence these scores~\cite{mallaband2024agreement, ramsdell2007predicting,shriberg1991reliability}. For instance, agreement is consistently lower for narrow transcription, which captures fine phonetic detail such as diacritics, than for broad transcription, which transcribes only phoneme-level distinctions. It is also generally lower for atypical than typical speech, for sounds that deviate from adult-like forms, for running speech over isolated words, and for certain segment classes such as vowels over consonants~\cite{shriberg1991reliability,stoel2001transcribing}. These findings suggest that a meaningful portion of our $42.1\%$ PER reflects ambiguity in the child speech signal and noise in the reference transcriptions, rather than recognition failures alone. Future work should quantify both human-human and human-machine agreement on the same speech samples, to establish the extent to which system errors exceed the variability present among human transcribers.

The second challenge is that naturalistic recordings are rife with competing signals: adult speech, sibling vocalizations, television, toys, and environmental noise. Even human-annotated boundaries are far from perfect: our data quality analysis found that 58 out of 200 sampled utterances contained untranscribed speech from other speakers. Rather than treating this as a preprocessing problem to be solved before recognition, we adopted an implicit approach: by training on large amounts of naturalistic data where only the target child's phonemes are labeled, the model learns, through fine-tuning, to attend to the relevant signal and suppress competing sources.  This is supported by the dramatic reduction in insertion rate from approximately $60\%$ for off-the-shelf baselines to $4.9\%$ for BabAR, as well as manual inspection of the predictions, which confirmed that BabAR tends to ignore adult speech present in the input segments. That said, more explicit approaches, such as incorporating a speaker enrollment module conditioned on the child's voice, could offer more robust handling of competing speakers. 

BabAR and TinyVox can be extended in several other directions, from age-specific models that account for the fundamentally different vocal characteristics of a 6-month-old and a 5-year-old, to phonotactic language models that bias decoding toward plausible phoneme sequences, though such models would likely need to be age-dependent, as the phonotactic structure of early babbling differs substantially from that of later, more complex speech. By making both TinyVox and BabAR publicly available, we hope to foster collaboration between speech technology and developmental science, enabling studies that have until now remained impractical at scale.

\section{Conclusion}

We introduced TinyVox, a cross-linguistic corpus compilation of over half a million phonetically transcribed child vocalizations, and BabAR, a phoneme recognition system that substantially outperforms existing baselines on child speech. Our experiments revealed that pretraining on multilingual child-centered daylong recordings and providing extended audio context during fine-tuning both improve performance, and that BabAR's errors tend to preserve broad phonetic categories. Applied to held-out naturalistic recordings of 44 infants, BabAR readily recovers established developmental trends without any manual annotation. By sharing both TinyVox and BabAR with the community, we hope to lower the barrier to large-scale phonetic analysis of child speech and foster tighter collaboration between speech technology and developmental science.

\section{Acknowledgments}

M.L. acknowledges funding from The Simons Foundation International (034070-00033). E.B. acknowledges funding from the National Institutes of Health (NIH DP5-OD019812). The authors gratefully acknowledge PhonBank (NIH-NICHD RO1-HD051698), and thank the data contributors whose corpora made this research possible. This work was performed using HPC resources from GENCI-IDRIS (2025-A0181011829).

\clearpage
\bibliographystyle{IEEEtran}
\bibliography{mybib}

\end{document}